\documentclass[aps,prl,twocolumn,showpacs,superscriptaddress,nofootinbib,groupedaddress]{revtex4} 

\usepackage{dcolumn}   % needed for some tables
\usepackage{pstricks}
\usepackage{color}
\usepackage{graphicx}
\usepackage{amsmath}
\usepackage{amssymb}
\usepackage[colorlinks=true,citecolor=darkred,urlcolor=darkred, pdfborder={0 0 0}]{hyperref}
\definecolor{darkred}{rgb}{0.6,0,0}

\definecolor{linkcolor}{rgb}{0,0,0.5}
\definecolor{darkred}{rgb}{0.6,0,0}

\graphicspath{{figs/}}

\def\gsim{\raise0.3ex\hbox{$\;>$\kern-0.75em\raise-1.1ex\hbox{$\sim\;$}}}
\def\lsim{\raise0.3ex\hbox{$\;<$\kern-0.75em\raise-1.1ex\hbox{$\sim\;$}}}

\def\beqn#1{\begin{equation}\label{#1}}
\def\eeqn{\end{equation}}

\def\beqa#1{\begin{eqnarray}\label{#1}}
\def\eeqa{\end{eqnarray}}

\newcommand{\fig}[1]{Fig.~(\ref{#1})}

\def\g{\gamma}
\def\g5{\gamma_5}
\def\21{SU(2) $\otimes$ U(1) }

\def\SM{SU(3)$_c$ $\otimes$ SU(2)$_L$ $\otimes$ U(1)$_Y$ }

\def\TrTrOne{ \rm{SU(3)$_c$ $\otimes$ SU(3)$_L$ $\otimes$ U(1)$_X$ }}

\def\znbb {$0\nu\beta\beta$ }

\def\Z2{$\mathcal{Z_2}$}

\def\vev#1{\left\langle #1\right\rangle}

\def\one{\ensuremath{\mathbf{1}}}

\def\three{\ensuremath{\mathbf{3}}}
\def\threeS{\ensuremath{\mathbf{3^*}}}

\newcommand{\gev}{\,\mathrm{GeV}}

\begin{document}

\title{Radiative neutrino mass in 3-3-1 scheme}

%\title{Calculable neutrino masses in a {\TrTrOne} electroweak gauge theory}

\author{Sofiane M. Boucenna}
\email{boucenna@ific.uv.es}
\affiliation{Instituto de F\'{\i}sica Corpuscular (CSIC-Universitat de Val\`{e}ncia), Apdo. 22085, E-46071 Valencia, Spain.}
\author{Stefano Morisi}
\email{stefano.morisi@gmail.com}
\affiliation{ DESY, Platanenallee 6, D-15735 Zeuthen, Germany.}
\author{Jos\'e W.F. Valle}
\email{valle@ific.uv.es}
\affiliation{Instituto de F\'{\i}sica Corpuscular (CSIC-Universitat de Val\`{e}ncia), Apdo. 22085, E-46071 Valencia, Spain.}
\date{\today}

\pacs{14.60.Pq, 12.60.Cn, 14.60.St, 14.70.Pw,  12.15.Ff   }
%%%%%%%%%%%%%%%%%%%

\begin{abstract}
\noindent

We propose a new radiative mechanism for neutrino mass generation
based on the \TrTrOne electroweak gauge group.  Lepton number is a
symmetry of the Yukawa sector but spontaneously broken in the gauge
sector. As a result light Majorana masses arise from neutral gauge
boson exchange at the one-loop level. In addition to the isosinglet
neutrinos which may be produced at the LHC through the extended gauge
boson \textit{portals}, the model contains new quarks which can also
lie at the TeV scale and provide a plethora of accessible collider
phenomena.

\end{abstract}

\maketitle

%%\section{Introduction}

The origin of neutrino mass and mixing, required in order to account
for neutrino oscillation data~\cite{Beringer:1900zz,Tortola:2012te},
poses one of the biggest challenges in particle physics.
While charged fermions must be Dirac particles, neutrinos are
generally expected to be Majorana fermions~\cite{Schechter:1980gr},
breaking lepton number and inducing neutrinoless double beta
decay~\cite{Schechter:1982bd}. While attractive, the idea that
neutrino mass is related to unification, encoded in the high-scale
seesaw
paradigm~\cite{Minkowski:1977sc,gell-mann:1980vs,yanagida:1979,mohapatra:1980ia,Schechter:1980gr}
falls short of covering the wealth of interesting neutrino mass
schemes. Indeed neutrino masses could well be a low-scale phenomenon,
both within the seesaw mechanism as well as other alternative
approaches~\cite{Boucenna:2014zba}.  This brings in substantial
freedom in model building, making the structure of the leptonic weak
interaction much richer than the CKM
matrix~\cite{cabibbo:1963yz,kobayashi:1973fv} characterizing the quark
sector, and opening the exciting possibility of probing the associated
neutrino mass messenger particles at collider
experiments~\cite{Deppisch:2013cya,AguilarSaavedra:2012fu,Das:2012ii}.

Lepton number symmetry provides an important theoretical guide in
neutrino mass modeling, depending on its fate different classes
of models can be envisaged.
For instance, lepton number can be conserved, leading to Dirac
neutrinos. Or it can be violated explicitly, since gauge singlet
Majorana masses can be added by hand in the \SM 
model~\cite{Schechter:1980gr}. Or it can be a spontaneously broken
global or gauged U(1) symmetry. The former defines the so-called
seesaw majoron schemes~\cite{chikashige:1980ui,Schechter:1982cv},
while the latter characterizes left-right symmetric electroweak
models~\cite{Senjanovic:1975rk}.
Another important challenge is the origin of the number of
families. We know that three different flavors exist, i.e. states with
the same gauge quantum numbers but different mass. But we do not know
why nature replicates, nor why the masses of the three generations of
Standard Model quarks and leptons are so different, nor why they mix
in the way they do (flavor problem).

In this paper we consider an alternative approach to neutrino mass
generation at accessible scales and "explaining" the number of
families. 
The model is based on the \TrTrOne (3-3-1) electroweak gauge structure
and is consistent only if the number of families equals the number of
quark colors~\cite{Singer:1980sw,valle:1983dk}, giving a reason for
having three species of fermions. This feature follows from gauge
anomaly cancellation and characterizes 331 models, including other
variants
e.g.~\cite{Foot:1994ym,PhysRevLett.69.2889,Pisano:1991ee,Hoang:1995vq,Tully:2000kk,Dong:2013wca}.

Our new mechanism for generating neutrino mass at the one-loop level
involves a neutral gauge-mediated lepton number-violating
interaction. All new particles, including the messengers associated to
neutrino mass generation can lie at the few TeV scale, hence
accessible to collider searches.\\[-.2cm]

%%\section{The model}

We start from the \TrTrOne gauge framework suggested
in~\cite{Singer:1980sw,valle:1983dk}. We concentrate on the
electroweak part of the model.  The left-handed leptons are assigned
to the anti-triplet representation of $SU(3)_L$
\begin{equation}
\psi_L^\ell=\left(
\begin{array}{c}
\ell^-\\
\nu_\ell\\
N_\ell^c
\end{array} \right)_L  \,,
\end{equation}
where $\ell=1,2,3 \equiv e,\,\mu,\,\tau$.  In addition to the new
two-component neutral fermions present in the lepton triplet
$N_L^c\equiv (N^c)_L\equiv (\nu_R)^c$ where $\psi^c =
C\overline{\psi}^T$ and $C$ is the charge conjugation matrix, we
introduce new sequential lepton-number-carrying gauge singlets $S=\{
S_1,S_2,S_3\}$
sequentially~\cite{mohapatra:1986bd,bernabeu:1987gr,Dittmar:1989yg,branco:1989bn,rius:1989gk}.
The matter content of the model is summarized in Tab.~(\ref{tab:content}).
\begin{table}[!h]
\begin{center}
\begin{tabular}{c|c|c|c|c|c|c|c|c|c|c}
\hline
 & $\psi_L^\ell$ & $\ell_R$ & $Q_L^{1,2}$  & $Q_L^3$ & $\hat{u}_{R}$ & $\hat{d}_{R}$ & $\,S\,$ & $\phi_1$ & $\phi_2$ & $\phi_3$ \\
\hline
\hline
$SU(3)_c$ & \one &\one &\three &\three &\three &\three &\one&\one&\one &\one \\
\hline
$SU(3)_L$ & \threeS & \one &\three & \threeS& \one & \one & \one& \threeS & \threeS& \threeS    \\
\hline
$U(1)_X$ & $-\frac{1}{3}$ & $-1$ & $0$ & $+\frac{1}{3}$ & $+\frac{2}{3}$ & $-\frac{1}{3}$ &  $0$& $+\frac{2}{3}$& $-\frac{1}{3}$& $-\frac{1}{3}$\\ 
\hline
$ \mathcal{L} $ & $-\frac{1}{3}$ & $-1$ & $-\frac{2}{3}$ & $+\frac{2}{3}$ & $0$ & $0$ &  $1$&$+\frac{2}{3}$ & $-\frac{4}{3}$&$+\frac{2}{3}$\\ 
\hline
% $ \mathbb{Z}_2^{\rm aux} $ & $+$ & $+$ & $+$ & $-$ & $+$ & $-$ &  $+$& $+$& $+$& $-$\\ 
\hline
\end{tabular}\caption{Matter content of the model, where
  $\hat{u}_R\equiv{(u_R,c_R,t_R,t'_R)}$ and $\hat{d}_R\equiv{(d_R,s_R,b_R,d'_R,s'_R)}$ (see text).} 
\label{tab:content}
\end{center}
\end{table}

\vskip -.4cm
With the above $\mathcal{L}$ assignment the electric charge and lepton
number are given in terms of the $U(1)_X$ generator $X$ and the
diagonal generators of the $SU(3)_L$ as
\beqa{eq:QL}
Q&=&T_3+\frac{1}{\sqrt{3}}T_8+X  \quad , \\
L&=&\frac{4}{\sqrt{3}} T_8+\mathcal{L}  \quad .
\eeqa

In order to spontaneously break the weak gauge symmetry, we introduce
three scalar anti-triplets $\phi_1\sim ({\bf 3}^*,+2/3)$ and
$\phi_{2,3} \sim ({\bf 3}^*,-1/3)$.  Note that the third component of
$\phi_3$ carries two units of lepton number. Following the notation of
\cite{valle:1983dk} we have the following vacuum expectation values
(VEVs)
\begin{equation}
\label{scalar3plets}
\vev{\phi_1}=\left[
\begin{array}{c}
k_1\\
0\\
0
\end{array}
\right],
\vev{\phi_2}=\left[
\begin{array}{c}
0\\
0\\
n_1
\end{array}
\right],
\vev{\phi_3}=\left[
\begin{array}{c}
0\\
k_2\\
n_2
\end{array}
\right]\,,
\end{equation}
where the $k_1$ and $k_2$ VEVs are at the electroweak scale and
correspond to the VEV of the $SU(2)_L\subset SU(3)_L$ doublets. The
VEVs $n_1$ and $n_2$ are isosinglet VEVs that characterize the
$SU(3)_L$ breaking scale.  Note that while $\phi_3$ takes VEV in both
electrically neutral directions, the second VEV of $\phi_2$ is
neglected, so that lepton number is broken only by $SU(2)_L$
singlets. This pattern gives the simplest consistent neutrino mass
spectrum, avoiding the linear seesaw
contribution~\cite{akhmedov:1995ip,Malinsky:2005bi}.\\[-.2cm]

%%%%%%%%%%%%%%%%%%%%%%%%%%%%%%%%%%%%%%%%%%%%%%%%%%%%%%%%%%%

The kinetic term for the scalar fields is 
\begin{equation}
\mathcal{L}_{\rm Kin} = 
\sum_{i}(D^\mu \phi_i)^\dagger(D_\mu \phi_i) \,.
\end{equation}
We define the covariant derivative as 
\begin{equation}
D_\mu \phi_{i}=\partial_\mu \phi_i +  i \frac{g_1}{\sqrt{2}} \, \mathbf{W}_\mu.\phi_i +  i \sqrt{2} g_2 X_i\, \mathbf{B}_\mu  \,\phi_i \,,
\end{equation}
where $X_i$ is the $U(1)_X$ charge of the $\phi_i$ scalar field,
$\mathbf{B}_\mu$ is the gauge boson of $U(1)_X$ and $\mathbf{W}_\mu
\equiv \sum_i W_\mu^i \lambda_i$ where $\lambda_i$ are the Gell-Mann
matrices. In matrix form one has
\begin{eqnarray}
\label{eq:W}
&& \mathbf{W}
=\begin{pmatrix}
W^3+\frac{1}{\sqrt{3}} W^8 & W_{12}^+ & W_{45}^+ \\
W_{12}^- & -W^3+\frac{1}{\sqrt{3}} W^8  & W^6-i W^7\\
W_{45}^-& W^6+i W^7 &-\frac{2}{\sqrt{3}} W^8 \\
\end{pmatrix}\,,\nonumber
\end{eqnarray}
where for simplicity we dropped the index $\mu$ from the gauge bosons
and defined the charged states as
\begin{equation}
W_{12}^\pm=\frac{1}{\sqrt{2}} (W^1\mp i W^2)\,,\quad
W_{45}^\pm=\frac{1}{\sqrt{2}} (W^4\mp i W^5)\,.
\end{equation}

There are in total nine electroweak gauge bosons, four of which are
charged ($W_{12}^\pm$ and $W_{45}^{\pm}$), giving the physical $W^\pm$ and $W^{\prime \pm}$, while five are electrically
neutral, namely $W^3,\,W^6,\,W^8,\,B$ giving the physical bosons
\vskip-3.5mm
\begin{equation}
\gamma \,,\quad Z\,,\quad Z'\,,\quad X
\end{equation}
and finally one  neutral boson, unmixed if CP is conserved 
\begin{equation}
W^7\equiv Y\,.
\end{equation}
\vskip-2.5mm
Assuming for simplicity $k_2\sim  k_1\ll n_1\sim n_2$ one finds
the gauge boson masses to leading order
\begin{eqnarray}
 m_{W}^2  &=&  g_1^2 \,(k_1^2+ \frac{k_2^2 n_1^2}{n_1^2+n_2^2}) \,,\nonumber \\
 m_{W'}^2  &=& g_1^2 (n_1^2+n_2^2) \,,\nonumber\\
m_Z^2\,\,    &=&  \frac{g_1^2(3 g_1^2+4 g_2^2)}{3 g_1^2+{g_2}^2}\, (k_1^2+\frac{k_2^2 n_1^2}{n_1^2+n_2^2})  \,, \\
m_{Z'}^2 &=& \frac{4}{9}(3 g_1^2+g_2^2) (n_1^2+n_2^2) \,, \nonumber\\
m_{X}^2  &=& m_{Y}^2 \, = \,   g_1^2 (n_1^2+n_2^2).\nonumber
%m_{Y}^2  &\approx& g_1^2 (n_1^2+n_2^2).
\end{eqnarray}

\noindent The ratio of the SM gauge bosons masses is given by
\begin{equation}
\frac{m_W^2}{m_Z^2}=\frac{3+\tan^2\theta_{331}}{3+4\tan^2\theta_{331}} \,,
% \quad,\quad \cos^2\theta= \frac{12 g_1^2+{g_2}^2}{24 g_1^2+8{g_2}^2}\,.
\end{equation}
where $\tan^2\theta_{331} \equiv g_2/g_1 $. The SM relation
$\cos^2\theta_W\equiv m_W^2/m_Z^2=0.76$
%which is equal in the SM to $\cos^2\theta_W=0.76$, 
implies that $\tan^2\theta_{331}= 0.57$.
The leptonic neutral current weak interaction $\mathcal{L}_{NC}$
contains
\begin{eqnarray}
\mathcal{L}_{NC}\, &&\supset 
\frac{g_1}{\sqrt{2}}\, \overline{\nu_L}\,\gamma_\mu\,\nu_L  W^3_\mu %\nonumber\\
% + \frac{2}{3}c_1\,\epsilon\, \, \overline{N^c_L}\,\gamma_\mu\,\nu_L\, Z 
% &&-\,
% c_2\, \overline{N^c_L}\,\gamma_\mu\,\nu_L\, Z^\prime\,\,
-\, \frac{g_1}{\sqrt{6}}\,\overline{\nu_L}\,\gamma_\mu\,\nu_L W^8_\mu  \, \nonumber\\
&&-\, \frac{g_1}{\sqrt{2}}\, \overline{N^c_L}\,\gamma_\mu\,\nu_L\, W^6_\mu + \frac{\sqrt{2} g_2}{3}\overline{\nu_L}\,\gamma_\mu\,\nu_L B_\mu
\,,
\label{nc}
\end{eqnarray}
The mixing of $W^6$ with $W^3, W^8$ and $B$ is proportional to the
small parameter given by
\begin{equation}
\label{eps}
\epsilon \, \sim \frac{k_2 n_2}{n_1^2+n_2^2}\ll 1\,.
\end{equation}
In addition to breaking the standard electroweak symmetry, this mixing
also violates lepton number by two units as can be readily seen
through its proportionality to $n_2$ which is the scale of lepton
number violation.  
The spontaneous symmetry breaking follows the pattern
\begin{equation}
SU(3)_L\otimes U(1)_X \xrightarrow{n_{1,2}}  SU(2)_L\otimes U(1)_Y 
 \xrightarrow{k_{1,2}} U(1)_Q. \nonumber
\end{equation}

 \vskip3.mm  %%%%%%%%%%%%%%%%%%%%%%%%%%%%%%%%%%%%%%%%%%%%%%%%%%%%%%%%%%%

\begin{figure}[t!]
\centering
\includegraphics[width=0.25\textwidth]{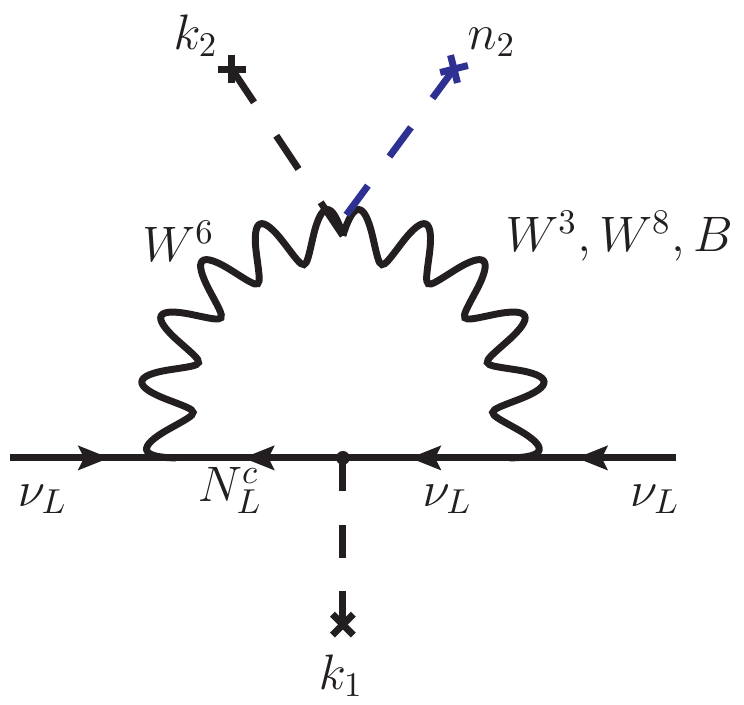}
\caption{Gauge boson exchange diagram for radiatively induced Majorana
  neutrino  mass in the flavor basis.}
\label{figd}
\end{figure}

Turning to the lepton sector, the Yukawa terms are
\begin{eqnarray}\label{lagY}
 \mathcal{L}_{\rm leptons} &=& \,
 y^{\ell}_{ij} \overline{\psi_L^i} \,l_R^j \,\phi_1 +  y^a_{ij} \,\psi_L^{i T} C^{-1}  \psi_L^j \,\phi_1 \\
&&+\, y^s_{ij} \,\overline{\psi_L^i} \, S^j \,\phi_2 + \mathrm{h.c.}\nonumber
\end{eqnarray}
where contraction of the flavor indices $i,j=1,2,3$ is assumed. Here
$y^{\ell}$ and $y^s$ are arbitrary matrices while ${y^a}$ is
antisymmetric.  The charged lepton mass matrix is just $M_\ell =y^\ell
\vev{\phi_1^0}$ and can be made diagonal in the usual way.  Note that,
thanks to an auxiliary parity symmetry, $\phi_3$ does not couple to
leptons.  The tree level neutrino mass matrix in the basis
($\nu_L,\,N^c,\,S$) is given by
\begin{equation}
\label{mnu1}
M_\nu = %\simeq
\left(
\begin{array}{ccc}
0& m_D &0\\
& 0 &M\\
& & 0
\end{array}
\right)\, , 
\end{equation}
where $m_D=k_1\,y^a$, 
%$M'=k_2\, y^s$ 
and $M=n_1\,y^s$.  Note that lepton number conservation forbids the
Majorana mass entry for $S$. 
We denote the corresponding eigenstates as $\nu_1,\,\nu_2,\,\nu_3$. 
The heavy states form Dirac pairs with masses ${M_D}_i$ ($i=1,2,3$)
given by
\begin{equation}\label{MD}
{M_D}_i = (\sqrt{{{m_D}\cdot {m_D}^T+M\cdot M^T}})_i\,,
\end{equation}
where the index $i$ in the r.h.s denotes the $i^{th}$ eigenvalue of
the
matrix~\cite{mohapatra:1986bd,bernabeu:1987gr,Dittmar:1989yg,branco:1989bn,rius:1989gk}.
On the other hand the state $\nu_1$ is massless because of lepton
number conservation in Eq.~(\ref{lagY}).
This holds at tree level.
However lepton number is broken spontaneously by $n_2 \neq 0$. This
induces light neutrino masses radiatively, as illustrated by the
diagram in Fig.~(\ref{figd}).
%%
%%%%%%%%%%%%%%%%%%%%%%%%%%%%%%
%%
In order to estimate the effective light neutrino mass scale which
results from Eq.~(\ref{mnu1}), we adopt for simplicity the one family
approximation (generalization to three is straightforward).
Indeed, the interplay of the intra-multiplet gauge boson exchange
connecting $\nu$ to $N^c$ with the gauge boson mixing implicit in
Eq.~(\ref{eps}) implies that lepton number is necessarily violated in
the neutral fermion sector. As a result the massless neutrino is not
protected and radiative corrections involving the gauge bosons will
yield a \textit{calculable} Majorana mass term as depicted in the
diagram of \fig{figd}.
To perform the corresponding estimate one goes to the mass basis. The
result is that the light neutrino $\nu_1$ gets a Majorana mass by
means of the exchange of the massive Dirac states $\nu_{2,3}$ and the
gauge bosons $Z$ and $Z^\prime$. Now we describe in more detail how
this works.

%%\vskip5.mm

Consider the $3\times 3$ mass matrix given in Eq.~(\ref{mnu1}).  This
matrix is diagonalized by an orthogonal matrix, given by a $1-3$
rotation followed by a $2-3$ maximal rotation
\begin{equation}
U_\nu=
\left(
\begin{array}{ccc}
c& -\frac{s}{\sqrt{2}} & \frac{s}{\sqrt{2}}\\
0& \frac{1}{\sqrt{2}} & \frac{1}{\sqrt{2}}\\
-s& -\frac{c}{\sqrt{2}} & \frac{c}{\sqrt{2}}
\end{array}
\right)\,
\end{equation}
\noindent where $c\approx 1$ and $s\equiv \beta \simeq m_D/M \ll 1$
characterizes the doublet-singlet mixing, which could be at most a few
percent due to universality constraints. Then the states
$\nu_L,N^c_L,S$ are related to the massive neutrino states
$\nu_1,\nu_2,\nu_3$ by (up to corrections of order $\epsilon$)
\begin{eqnarray}\label{numassbasis}
\nu_L&\simeq&  \nu_1-\beta\,\nu_2+\beta\,\nu_3\,;\nonumber\\
N^c_L&\simeq&\frac{1}{\sqrt{2}}(\nu_2+\nu_3)\,;\\
S&\simeq&-\beta\,\nu_1  + \frac{1}{\sqrt{2}}(-\nu_2+\nu_3)\nonumber \,.
\end{eqnarray}
%
% From Eq.~(\ref{nc}) and (\ref{numassbasis}), the gauge interaction
% terms relevant in inducing the light neutrino mass, namely
% \begin{equation}
%  \overline{\nu_L}\,\gamma_\mu\, \nu_L\, Z^{(\prime)}\,,\qquad
% \epsilon\,\overline{N^c}\,\gamma_\mu\, \nu_L\, Z^{(\prime)}\,,
% \end{equation}
% produce the couplings
% % 
% \begin{equation}
% \beta\,\overline{\nu_1}\,\gamma_\mu\, \nu_{2,3}\, Z^{(\prime)}\,,\qquad
% \frac{\epsilon}{\sqrt{2}} \, \overline{\nu_1}\,\gamma_\mu\, \nu_{2,3}\, Z^{(\prime)}\,,
% \end{equation}
% respectively, where $Z^{(\prime)}$ means either $Z$ or $Z^\prime$.
% %
Although both $Z$ and $Z^\prime$ enter in the loop, the main
contribution is from $Z^\prime$ exchange, estimated from
Eq.~(\ref{numassbasis}) and the diagonalization of the gauge bosons to
be
\begin{equation}\label{mllrad}
m_{\nu_{\rm light}}\simeq \frac{g^2  \, \epsilon\, \beta}{16 \pi^2}  M_D \frac{m_{Z^\prime}^2}{M_D^2+m_{Z^\prime}^2}\log \frac{m_{Z^\prime}^2}{M_D^2}\,,
\end{equation}
where $g$ is a simple function of gauge coupling constrants $g_1$ and
$g_2$.  Note that the contribution proportional to $\epsilon^2$ and
$\beta^2$ vanish as expected.

\begin{figure}[h!]
\centering
\includegraphics[width=0.35\textwidth]{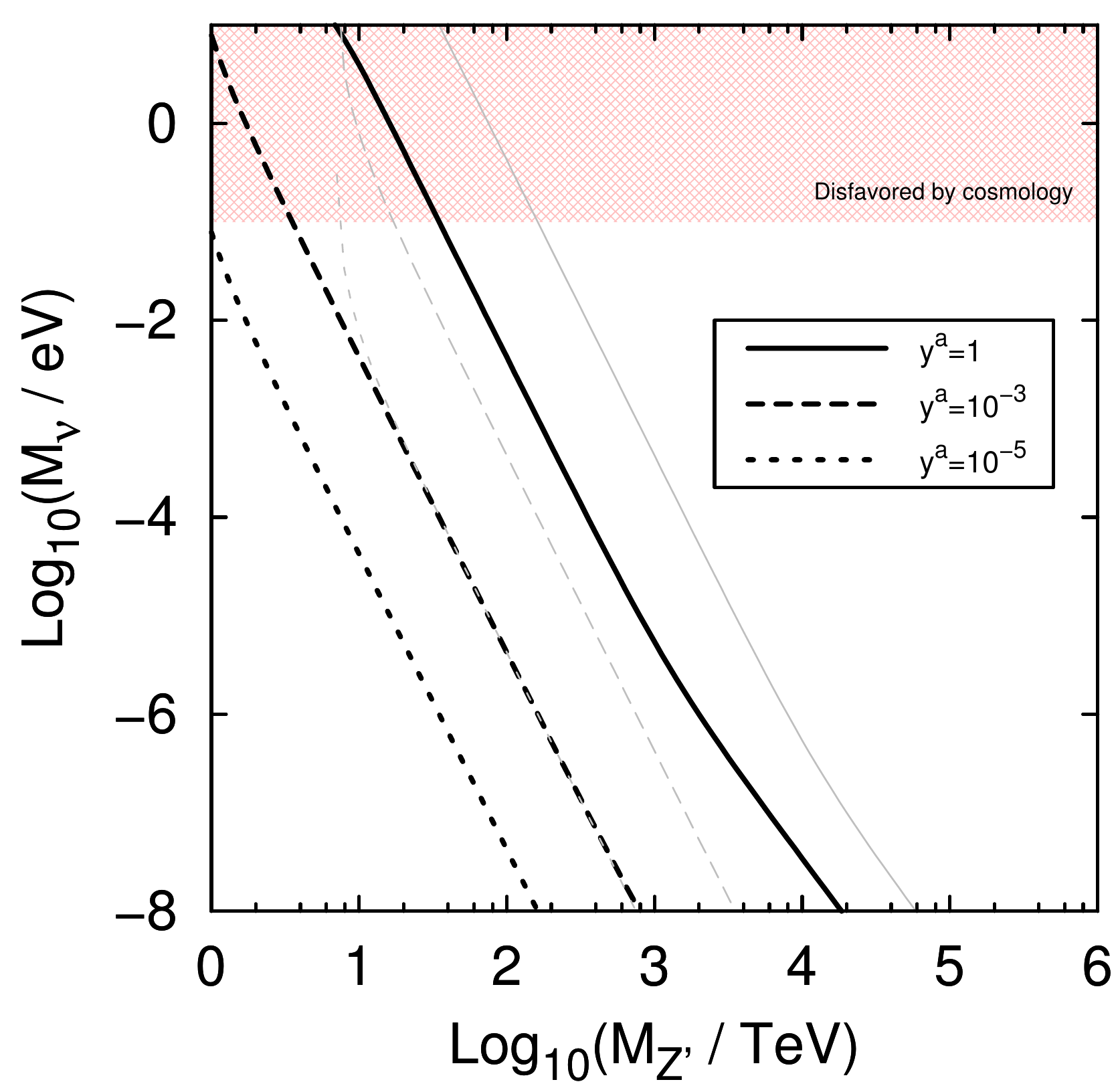}
\caption{Neutrino mass versus $Z^\prime$ scale for various values of
  the Dirac mass parameter $M_D$. Solid, dashed and dot-dashed lines
  correspond to $y^a=1$, $10^{-3}$ and $10^{-5}$ respectively.  $g_1=0.6$
  and $k_2=90 \gev$ and the scale of the new colored states ($n_2$) is fixed at 1 TeV (thick lines) and 10 TeV (thin lines).}
\label{figmnumz}
\end{figure}
Fig.~(\ref{figmnumz}) shows the correlation between the light neutrino
mass scale and the $Z^\prime$ mass for various values of the Dirac
mass $M_D$, parametrized by the Yukawa coupling $y^a$. For
definiteness we fix the $n_2$ VEV, responsible for the masses of the
new iso-singlet colored states at 1 TeV and 10 TeV.  Increasing $n_2$ would push
up the $Z^\prime$ mass and, assuming Yukawas of order one, would
increase the exotic (primed) quark masses.\\[-.3cm]

%%%%%%%%%%%%%%%%%%%%%%%%%%%%%%%%%%%%%%%%

This brings us to the discussion of the quark sector. The two first
generations $Q_L^1$ and $Q_L^2$ transform as triplets of $SU(3)_L$
whereas the third one $Q_L^3$ is anti-triplet:
\begin{equation}
%\sim ({\bf 3}^*,-1/3)\,
%
Q_L^1=\left(
\begin{array}{c}
u\\
d\\
d'
\end{array}
\right)_L ,\,
Q_L^2=\left(
\begin{array}{c}
c\\
s\\
s'
\end{array}
\right)_L ,\,
Q_L^3=\left(
\begin{array}{c}
b\\
t\\
t'
\end{array}
\right)_L \, .
\end{equation}

All right-handed states are $SU(3)_L$ singlets.  In order to pair up
the new left handed fields of the quark sector, we introduce extra
right handed fields $d'_R$, $s'_R$ and $t'_R$ that are singlets of
$SU(3)_L$.
Note that the axial anomaly cancels in this model because we have an
equal number of triplets and anti-triplets and the sum of the electric
charges on all the fermions vanishes \cite{Georgi:1972bb}.

The Yukawa Lagrangian of the quark sector is
% \vskip1mm
\begin{eqnarray}\label{lagQ}
\mathcal{L}_{\rm quarks}&=&  
y^{u}_{\alpha,i}\, \overline{Q_L^{\alpha}} \,\hat{u}_R^i \,\phi_1^* +  
y^{u}_{3,i}\, \overline{Q_L^{3}} \,\hat{u}_R^i \,\phi_3 \nonumber \\
&+& 
y^{d}_{3,i} \,\overline{Q_L^{3}} \,\hat{d}_R^i \,\phi_1 + %\\
y^{d}_{\alpha,i} \, \overline{Q_L^{\alpha}} \,\hat{d}_R^i \,\phi_3^* 
 + \mathrm{h.c.}
\end{eqnarray}
and $\alpha$=1,2. Contractions over $i$ and $\alpha$ indices are
assumed.  The up ($y^u$) and down ($y^d$) quark mass matrices are
respectively $3\times 4$ and $3\times 5$. Note that we made use of the
same auxiliary parity symmetry to charge $Q_L^3$ and $\hat{d}_R$.
After spontaneous symmetry breaking, the top mass is proportional to
$k_2$ while the bottom mass is proportional to $k_1$.  Hence $k_1$ and
$k_2$ determine the electroweak scale. On the other hand, the masses
of the extra quarks $d',\, s',\, t'$ are proportional to $n_1$ and
$n_2$ which must be of the order of TeV or greater in order to escape
detection at the LHC.  Here we note a novel feature of this model,
namely, that while the extra quarks have standard electric charges,
they carry two units of lepton number, relating them directly to the
lepton sector.

%%%%%%%%%%%%%%%%% 

In summary, we have proposed a new mechanism to generate neutrino mass
based on the $SU(3)_L \otimes U(1)_X$ gauge symmetry. At tree level
neutrinos are massless because of lepton number conservation. Gauge
interactions violate lepton number and lead to a Majorana mass term
for light neutrinos at one-loop level.
In contrast to most neutrino mass generation schemes, such as the
seesaw mechanism, where the neutrino mass comes from Yukawa couplings,
here it arises directly from gauge boson exchange as seen in
Fig.~(\ref{figd}) and Eq.~(\ref{mllrad}).  All neutrino species are
massive, and their splittings and mixing parameters can be fitted to
the oscillation data. The further imposition of genuine flavor
symmetries would bring in predictions for oscillation parameters and
possibly \znbb though we leave this for a separate investigation. Note
that the messenger particles responsible for neutrino mass can be
directly produced at the LHC: the $Z^\prime$ would de produced in
Drell-Yan process and provide a "portal" to access the isosinglet
neutral leptons~\cite{Deppisch:2013cya}. Moreover, if light enough,
the new exotic colored states would also be produced at the LHC and
induce gauge-mediated flavor-changing neutral currents, e.g. $b\to
s\mu^+\mu^-$~\cite{Buras:2013dea} providing a double test. These
issues will be taken up elsewhere.\\[-.2cm]

%%%%%%%%%%%%%%%%%

We thank Martin Hirsch and Renato Fonseca for useful discussions.  The
work of SB and JV was supported by MINECO grants FPA2011-22975 and
Multidark Consolider CSD2009-00064. SM thanks DFG grant WI 2639/4-1.

%%\bibliography{merged,newrefs,refs}

\end{document}